\documentclass[runningheads]{llncs}
\usepackage[T1]{fontenc}
\usepackage{graphicx}
\usepackage{subcaption}
\usepackage{amsmath}

\begin{document}
\title{Comparative Analysis of Hash-based Malware Clustering via K-Means}
%
%
\author{Aink Acrie Soe Thein\inst{1} \and
Nikolaos Pitropakis\inst{2,1}\orcidID{0000-0002-3392-9970} \and
Pavlos Papadopoulos\inst{1}\orcidID{0000-0001-5927-6026} \and
Sam Grierson\inst{1}\orcidID{0000-0002-3625-6337} \and
Sana Ullah Jan\inst{1}\orcidID{0000-0003-3950-4719}
}
\authorrunning{Soe Thei et al.}
%
\institute{Edinburgh Napier University, Scotland, UK \email{40542531@live.napier.ac.uk, p.papadopoulos@napier.ac.uk, S.Grierson2@napier.ac.uk} \and
Department of Information Technology, The American College of Greece, Athens, Greece
\email{npitropakis@acg.edu} \\}
\maketitle              
\vspace{-30pt}
\begin{abstract}
With the adoption of multiple digital devices in everyday life, the cyber-attack surface has increased. Adversaries are continuously exploring new avenues to exploit them and deploy malware. On the other hand, detection approaches typically employ hashing-based algorithms such as SSDeep, TLSH, and IMPHash to capture structural and behavioural similarities among binaries. This work focuses on the analysis and evaluation of these techniques for clustering malware samples using the K-means algorithm. More specifically, we experimented with established malware families and traits and found that TLSH and IMPHash produce more distinct, semantically meaningful clusters, whereas SSDeep is more efficient for broader classification tasks. The findings of this work can guide the development of more robust threat-detection mechanisms and adaptive security mechanisms.

\vspace{-10pt}
\end{abstract}

\vspace{-30pt}
\section{Introduction}
\vspace{-12pt}



Malware is software that infiltrates, damages, or takes control of systems without user consent. The scale of malware campaigns over the years has been shocking. Security vendors report around 350,000 new malicious programmes and potentially unwanted applications every day, and in 2020 alone, around 897 million malware instances have been discovered \cite{pachhala_comprehensive_2021}. In 2024, as reported by CrowdStrike \cite{CrowdStrikeThreatReport2025}, interactive intrusions accelerated dramatically: the average "breakout" time fell to 48 minutes (down from 62 minutes in 2023), with the fastest lateral move completed in just 51 seconds. Hands-on-keyboard, "malware-free" attacks accounted for 79\% of detections (up from 40\% in 2019), while vishing campaigns exploded by 442\% between the first and second half of the year. Moreover, 52\% of exploited vulnerabilities were used for initial access, and access broker advertisements rose 50\% year-over-year. Adversaries also began leveraging generative AI, enabling campaigns, such as \textit{FAMOUS CHOLLIMA}’s use of AI-crafted fake IT job candidates and state-aligned disinformation operations. These figures illustrate the rapid evolution of malware and the importance of catching it early.



Over the years, researchers and practitioners have developed a range of detection methods such as signature matching, heuristic analysis, behavioural monitoring, sandboxing, and hybrid systems, to keep up with emerging threats. Many signature-based tools rely on hashes to identify known malware: traditional cryptographic hashes like MD5, SHA-1, and SHA-256, as well as fuzzy-hash techniques (e.g. SSDeep, TLSH) that can spot variations of the same code \cite{lee_lightweight_2017}. However, malware continually evolves using polymorphic and metamorphic techniques that outpace traditional signature‐based detection methods \cite{badhwar_polymorphic_2021}.


In this work, we utilised SSDeep, TLSH, and IMPHash methods and fed their outputs into a K-Means clustering pipeline. The purpose of this work is to first evaluate the impact of clusters on various malware families and, secondly, to inform future work on improving classification techniques and threat hunting. Our contributions can be seen as follows: 
\vspace{-6pt}
\begin{itemize}
 \item Clustering of malware samples using the K-means algorithm, focusing on the accuracy of each hashing method against common malware families and behaviours.
\item Empirical analysis of the methods focusing on efficiency in terms of computational expenses and scalability.
\item Robustness investigation on changing malware characteristics and dataset fluctuations.
\end{itemize}
\vspace{-6pt}
The rest of the paper is organised as follows. Section \ref{related} reviews related work on hashing and malware detection. Section \ref{methodology} describes our testbed setup and data collection. Section \ref{results} presents and evaluates the clustering outcomes. Finally, Section \ref{conclusion} concludes and outlines directions for future work.

\vspace{-10pt}
\section{Background and Related Work}
\label{related}
\vspace{-10pt}

Once a malware instance is deployed, it can exfiltrate sensitive data, encrypt or corrupt data files, hijack the computational resources of the ``victim'' machine for unauthorised activities such as cryptocurrency mining, or launch attacks against others' machines, all that without the user's consent \cite{pachhala_comprehensive_2021}. For the detection of malware instances, various defensive mechanisms have been developed, such as \textit{Static analysis} and \textit{Signature-based detection}. Static analysis of the contents of the malware binary includes analysis of the source code, its structure, metadata, and embedded strings, all without the actual execution of the malware, so firewalls can match them against known malware signature or pattern rules \cite{Molina-Coronado2023-ok}. Similarly, signature-based detection involves comparing the malware's hash against public vulnerability databases of known malware families \cite{Khan2024-zz} to enable rapid detection of known threats \cite{Khan2024-zz}. However, a significant disadvantage of signature-based detection is that if the malware binary has a slight modification, then the produced hash will be different, hence, the firewall will not be able to detect its signature. More advanced deep inspection methods involve Heuristic-based Malware Static analysis (HMST) frameworks that employ a six-stage process to uncover suspicious characteristics beyond simple pattern matching, such as the verification of the hash, PE structure analysis, packer detection, measurement of entropy, antivirus scans via common platforms (such as VirusTotal \cite{virustotal}), and string analysis. Typical tools for disassembling malware binaries include Hex-Rays IDA Pro and OllyDbg, among others, which are also used to unpack and deobfuscate binaries and reveal hidden calls, paths, and attack payloads  \cite{hexraysRaysStateoftheart,olly}.

In situations where static code analysis techniques fall short, \text{Dynamic Analysis} techniques have been developed. Those include the careful monitoring of the malware binary upon its actual execution. In order to protect the examination machine from malware infection, sandboxed environments are used (such as virtual machines and emulators) to mimic a ``real'' environment and observe runtime malware behaviours such as external calls to adversary IP addresses, read or write of files, modifications in the system's registry \cite{Sihwail2018ASO}. A more advanced technique in this category is \textit{Behavioural Analysis} of binaries, which can classify them as benign or malicious while accounting for potential evasion techniques, external communication patterns, and persistence strategies \cite{ozturk2024dynamic}. This subcategory also includes detection using deep learning models, such as Convolutional Neural Networks (CNNs) combined with Long Short-Term Memory (LSTM) to capture deeper characteristics of malware, achieving excellent results (96\% accuracy) in zero-day malware detection \cite{Karat2024-mw}. Another example is the \textit{anomaly-based detection}, which first creates a ``profile'' of normal system or network activity during its training phase and then tests against live data to flag any deviations from that profile, making it particularly effective against novel threats \cite{Sharma2023-rs}.

However, dynamic analysis is resource-intensive, demanding CPU, memory, and time, which limits scalability. Sophisticated malware can recognise virtualised environments and alter its behavior to evade observation \cite{Aboaoja2023-lc}. Recognising the trade-offs of pure static or dynamic methods, \textit{hybrid systems} combine quick signature checks with deeper behavioral monitoring to balance throughput and depth \cite{Geng2023-rj}. Recent approaches advocate fusing signature-based and behavior-based detection into unified pipelines, leveraging the low false positives of static hashes alongside the zero-day coverage of anomaly detectors \cite{Kwon2022-bd}. Despite these advances, polymorphic and metamorphic malware continue to challenge conventional defenses by automatically rewriting or encrypting their code on each deployment. 
To overcome the brittleness of exact signatures and the cost of full behavioral emulation, \textit{perceptual} or \textit{similarity hashing} techniques generate compact fingerprints that tolerate code and structural variations while preserving core likeness. By clustering or linking these fuzzy hashes, security systems can identify related malware variants and detect novel threats with low overhead. In this work, we evaluate three representative hashing schemes, SSDeep, TLSH, and IMPHash, as features for unsupervised K-Means clustering in static malware analysis.

\vspace{-15pt}

\subsection{Hashing}
\vspace{-10pt}

Hashing transforms input data 
into a fixed-length string of characters using a mathematical function \cite{sansOverviewCryptographic}. In cybersecurity, hashes serve multiple roles, including integrity checks, password storage, and signature-based malware detection. Traditional cryptographic hashes (MD5, SHA-1, SHA-256) detect exact matches but break if even a single bit changes. To address this brittle behaviour, \textit{fuzzy} or \textit{similarity} hashing was introduced. These algorithms produce digests that tolerate both minor and significant file modifications, making them ideal for spotting variants of the same malware \cite{Lee2017-km}. Broad categories include Context-Triggered Piecewise Hashes (CTPH), multi-resolution similarity hashes, feature-extraction-based methods, and locality-sensitive hashes \cite{Lee2017-km,Sharma2023-rs}.



\textbf{SSDeep} is a CTPH method adapted from the \textit{spamsum} tool for email signatures \cite{kornblum_identifying_2006}. It divides a file into context-driven segments of varying size, computes a rolling hash on each segment, and concatenates the results. SSDeep outputs a similarity score from 0 to 100: 100 indicates identical content, while 0 means no detectable overlap. Because each segment hashes independently, an attacker must alter large portions of the file before the SSDeep score drops significantly. 



\textbf{SDHash} differs from other methods, since it builds a similarity digest by selecting rare, high-entropy byte features from a sample, hashing them using the SHA-1 algorithm, and recording them in a compact Bloom filter. Two digests are compared by estimating how many features they share (mitigating Bloom collisions) and returning a 0–100 similarity score, with higher meaning that they are more similar \cite{breitinger_frash_2013}. Extensive benchmarks show SDHash consistently outperforms SSDeep in recall and precision, at the cost of slower computation due to its reliance on SHA-1 \cite{roussev_evaluation_2011}.


\textbf{Trend Micro’s Locality-Sensitive Hash (TLSH)} generates a digest only if the input is at least 50 bytes long. Unlike SSDeep and SDHash 
, TLSH defines 0 as an exact match; larger values indicate greater dissimilarity \cite{oliver_tlsh_2013}. Tests confirm that TLSH effectively groups files with small edits, offering balanced runtime, low false-positive rates, and resistance to simple evasion tactics.


\textbf{Import hashing (IMPHash)} targets Windows Portable Executable (PE) files by recording the sequence of imported libraries (such as .DLL, .EXE, .SYS) and hashing that sequence \cite{naik_fuzzy-import_2020}. Because import order often reflects a developer’s coding style or a toolchain’s behaviour, matching IMPHashes can link samples to the same author or toolkit. IMPHash outputs a binary match (“yes”/“no”) rather than a graded score, so it is fast but fragile: reordering imports, packing, or obfuscation easily thwarts it, and it only applies to Windows binaries \cite{Naik2019}.



As seen previously, similarity-hashing methods have their strengths and weaknesses, and no method can detect evasion techniques effectively. In this work, we integrate SSDeep, SDHash (via its Bloom-filter digest), TLSH, and IMPHash into a K-means clustering pipeline and compare their performance for static malware analysis.

\vspace{-12pt}
\section{Methodology and Implementation}
\label{methodology}
\vspace{-10pt}


\subsection{Dataset Selection}
\vspace{-8pt}



Using labelled datasets is critical for evaluating malware clustering techniques. However, there is no fundamentally better way to assemble them. In this work, we investigated the following datasets:

\vspace{-8pt}

\begin{itemize}
	\item \textbf{BODMAS} \cite{yang_bodmas_2021}. This dataset was created from 134,435 Windows PE samples (57,293 malware and 77,142 benign samples), gathered from August 2019 to September 2020. BODMAS provides labels for the samples (malware/benign) alongside the ``families'' they belong to, addressing limitations in earlier PE-based classification datasets.
	\item \textbf{EMBER} \cite{anderson_ember_2018}. This dataset was collected between 2017 and 2018, and contains SHA‐256 hashes and binary labels (malicious/benign). The main weakness compared to the previous one is that it does not include malware family or type labels, which compromises further analysis as threats evolve.
	\item \textbf{APT Malware Dataset} \cite{githubCyberresearchAPTMalware}: Curated by Cyber Research APT Malware and used by Kida and Olukoya \cite{kida2022nation}, it includes 3,500 samples from 12 APT groups (allegedly sponsored by five nation‐states). It provides MD5, SHA‐1, SHA‐256 hashes plus metadata fields (country, apt‐group, family, Status).
\end{itemize}
\vspace{-8pt}

While SHA‐256 proved valuable for cross‐referencing samples on platforms such as VirusTotal \cite{virustotal} and MalwareBazaar \cite{abuseMalwareBazaar}, none of these datasets simultaneously offered a large sample size, rich family labels, and diverse malware types tailored to hashing experiments. Therefore, we assembled a new custom dataset which underpins our subsequent clustering and evaluation, ensuring that SSDeep, TLSH, and IMPHash are tested on a representative, well‐labelled collection of modern malware. The process is described as follows:
\vspace{-11pt}

\begin{itemize}
	\item \textbf{Sample sourcing}: We downloaded only PE‐format executables from Malware Bazaar, ensuring coverage of major families (e.g., Trojans, worms, ransomware).
	\item \textbf{Curation}: Samples were selected to balance family diversity and maintain realistic distributions, mirroring the variety encountered in the wild.
\end{itemize}
\vspace{-16pt}

\vspace{-10pt}

\subsection{Machine Learning Algorithm and Framework}
\vspace{-10pt}


One of the core tasks of clustering in unsupervised learning is to partition a dataset into subsets whose members are more similar to each other than to those in other subsets. While hierarchical clustering builds a tree of nested clusters by iteratively merging or splitting based on pairwise distances, K-Means directly partitions the data into a fixed number of clusters \textit{K}, which must be specified in advance.
The K-Means algorithm proceeds by first selecting \textit{K} initial centroids (often at random). In each iteration, every sample is assigned to the nearest centroid using Euclidean distance, and then each centroid is recomputed as the mean of the samples assigned to it. These two steps repeat until cluster memberships stabilise or a maximum number of iterations is reached. Because each update involves only simple arithmetic on vectors, K-Means scales roughly linearly with both the number of samples and the number of clusters, making it exceptionally well-suited for large-scale malware datasets.




We built the clustering pipeline in Python and used scikit-learn for the KMeans algorithm and basic pre-processing. For each hashing method, we first computed a digest for every sample. We then converted each digest into a fixed-length vector, such as that for SSDeep and TLSH by extracting numeric fields from their signatures, and for IMPHash by one-hot encoding the presence of known import-sequence hashes. We scaled each feature set and ran the same K-Means procedure on them independently (same initialisation, stopping rules, and number of restarts). This allowed us to compare how well the clusters match known malware families and behavioural labels, while keeping the clustering method constant.

\vspace{-20pt}

\subsection{Implementation}
\vspace{-10pt}





\begin{figure}[h!]
\centering
 \includegraphics[width=0.7\textwidth]{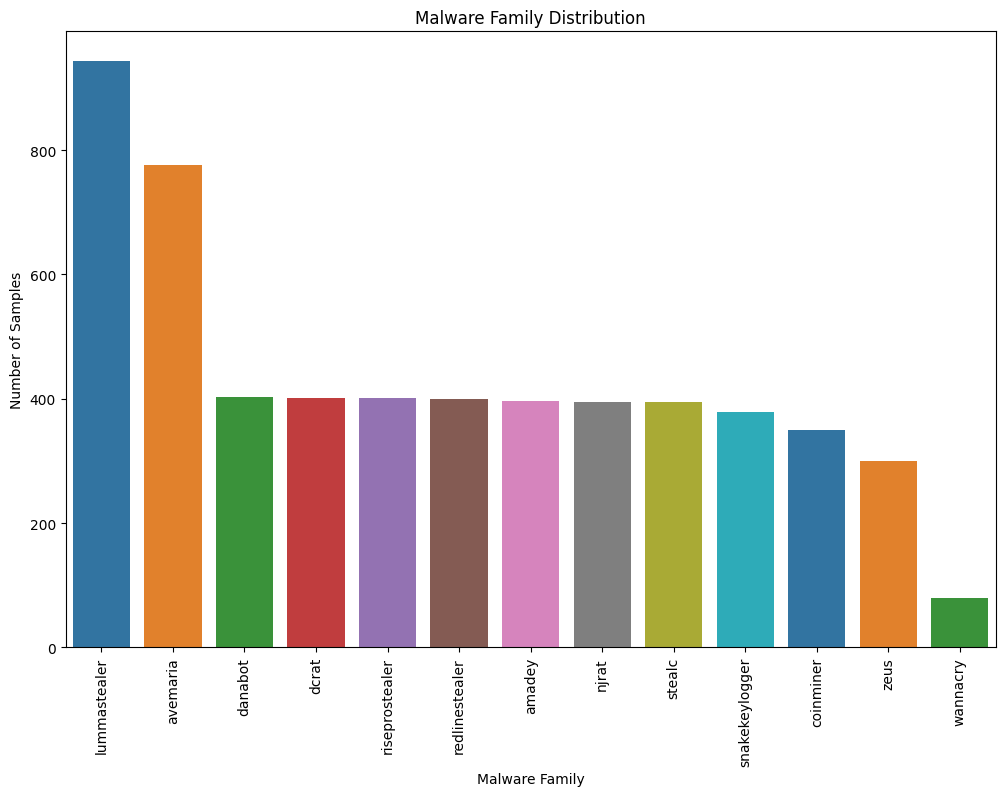}
 \caption{Malware Family Distribution Samples}
 \label{fig:distribution}
 \vspace{-20pt}
 \end{figure}

We verified dataset integrity by cross-checking each sample's SHA-256 hash against VirusTotal, MalwareBazaar, and VX Underground (free and commercial tiers). Because MalwareBazaar was the most consistent, we treated it as the primary source of information. For every sample, we recorded the SHA-256, first-seen and last-seen timestamps, and the malware family label in a CSV. To capture short-term shifts, we focused on PE samples collected from January-March 2024, and discovered, for example, a January rise in AgentTesla followed by the emergence of Mirai (see Figure \ref{fig:distribution}).

\vspace{-20pt}

We then enriched each sample with SSDeep, TLSH, and IMPHash digests using a Python pipeline. Standard libraries handled file traversal, CSV export, and SHA-256 computation (\texttt{hashlib}), while \textit{ssdeep}, \textit{tlsh}, and \textit{pefile} provided fuzzy hashing and PE import parsing. We merged all digests with the metadata on the SHA-256 key to produce a single unified table. To ensure complete features, we excluded files smaller than 50 bytes and samples without a valid import table, retaining only records containing SSDeep, TLSH, and IMPHash values. For clustering, we adopted Euclidean distance (the default in scikit‐learn’s KMeans) as our primary metric \cite{James2023}. The Euclidean distance between two \(n\)-dimensional vectors \(p\) and \(q\) is defined as in Equation \ref{eq:euc}. 
This metric reflects straight-line distance in feature space, aligning with the continuous similarity scores of SSDeep and TLSH. Since IMPHash produces binary matches, we used the Jaccard similarity, which suits variable-length import sequences. The Jaccard index for sets \(X\) and \(Y\) is defined in Equation \ref{eq:jacc}.
\begin{minipage}{.5\linewidth}
\small
\begin{equation}
 d(p, q) = \sqrt{\sum_{i=1}^{n} (q_i - p_i)^2}
 \label{eq:euc}
 \end{equation}
\end{minipage}%
\begin{minipage}{.5\linewidth}
\begin{equation}
 J(X,Y) = \frac{|X \cap Y|}{|X \cup Y|}
 \label{eq:jacc}
\end{equation}
\end{minipage}
\vspace{-10pt}

\section{Results and Evaluation}
\label{results}
\vspace{-10pt}

\begin{figure}[t!]
  \centering
  \makebox[\textwidth][c]{%
    \subfloat[Result of SSDeep cluster.]%
      {\includegraphics[width=0.50\textwidth]{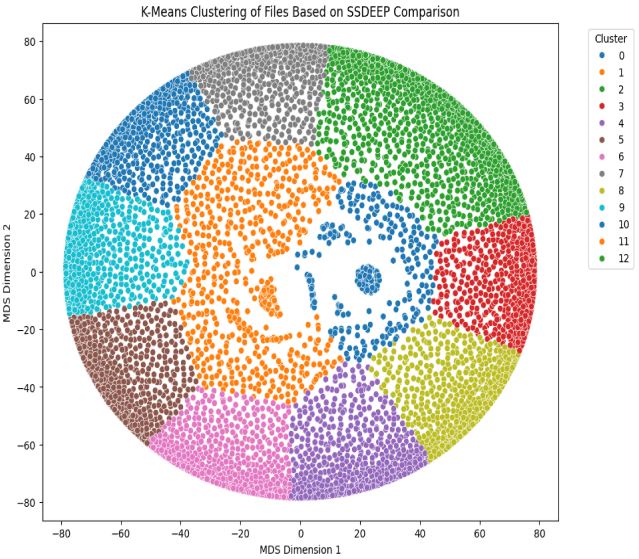}}\hspace{0.8em}%
    \subfloat[Result of TLSH cluster.]%
      {\includegraphics[width=0.455\textwidth]{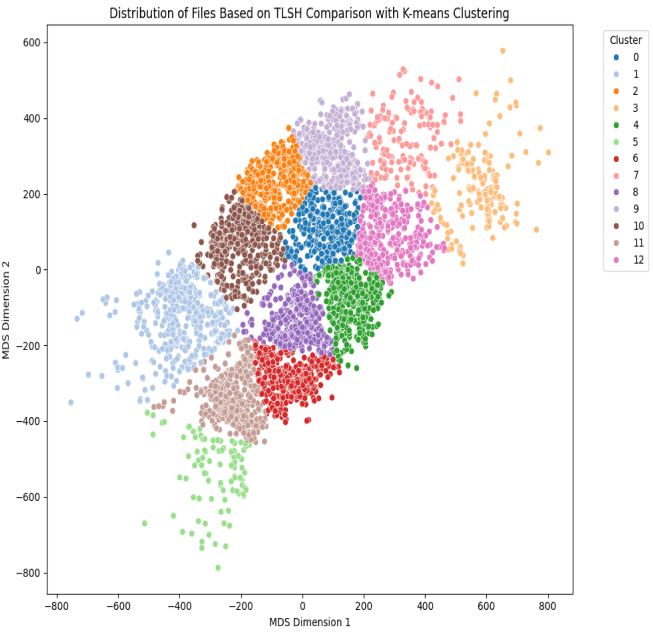}}\hspace{0.8em}%
    \subfloat[Result of Imphash cluster.]%
      {\includegraphics[width=0.46\textwidth]{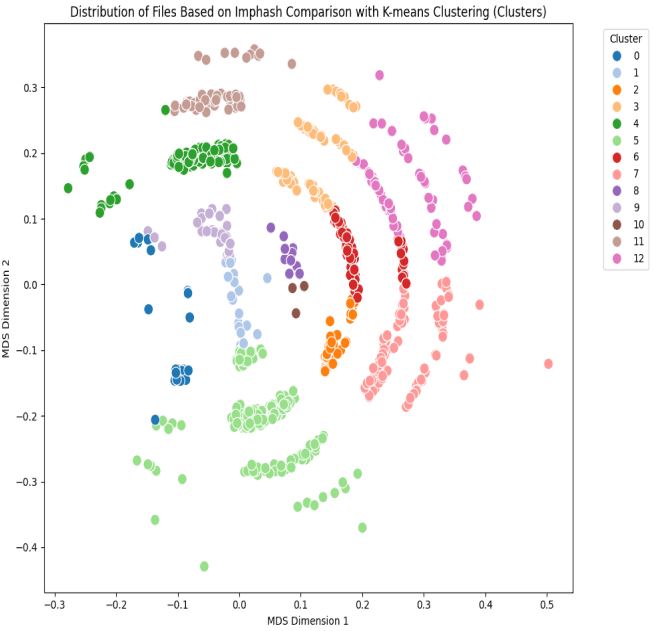}}%
  }
  \caption{Results of methods as clusters.}
  \label{fig:img_group_clusters}
  \vspace{-20pt}

\end{figure}

Here, we assess how well K-Means separates malware samples when using each hash type as a feature. Figure \ref{fig:img_group_clusters} shows the raw cluster layouts for SSDeep, TLSH, and IMPHash, respectively. With SSDeep (Fig. \ref{fig:img_group_clusters} (a)), the points form a large homogeneous cloud, showing that SSDeep captures general similarities but lacks the precision needed to split samples into tight groups. TLSH’s plot (Fig. \ref{fig:img_group_clusters} (b)) displays a clearer diagonal gradient, reflecting its ability to register small, incremental changes across files and to produce more distinct cluster boundaries. 
IMPHash (Fig. \ref{fig:img_group_clusters} (c)) forms compact, variably dense clusters, grouping samples with similar import sequences and suggesting shared family or author traits.







\begin{figure}[t!]
  \centering
  \makebox[\textwidth][c]{%
    \subfloat[Result of TLSH with family.]%
      {\includegraphics[width=0.5\textwidth]{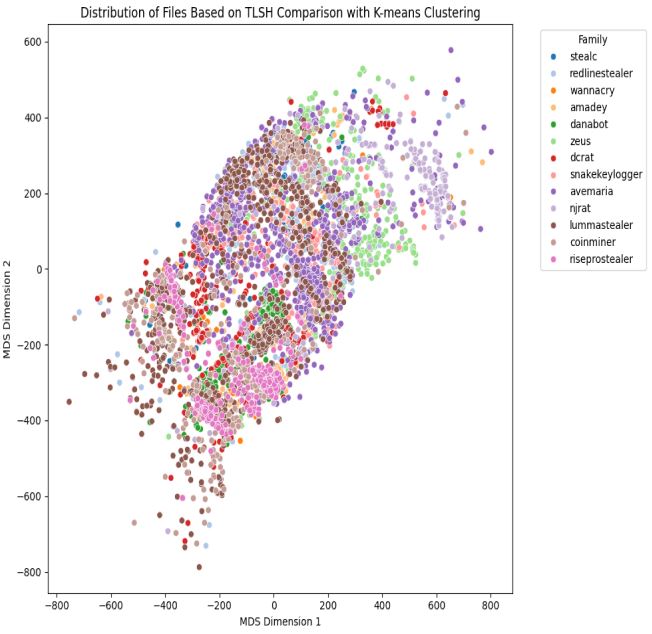}}\hspace{0.8em}%
    \subfloat[Result of SSDeep with family.]%
      {\includegraphics[width=0.5\textwidth]{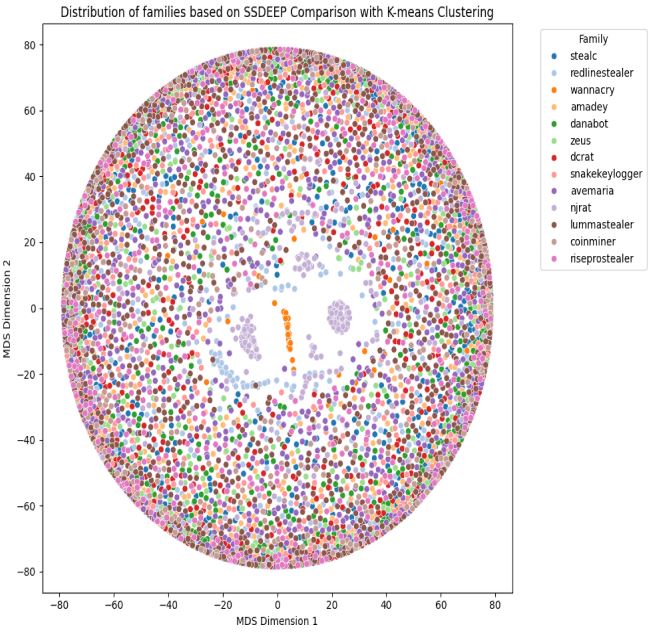}}\hspace{0.8em}%
    \subfloat[Result of ImpHash with family.]%
      {\includegraphics[width=0.5\textwidth]{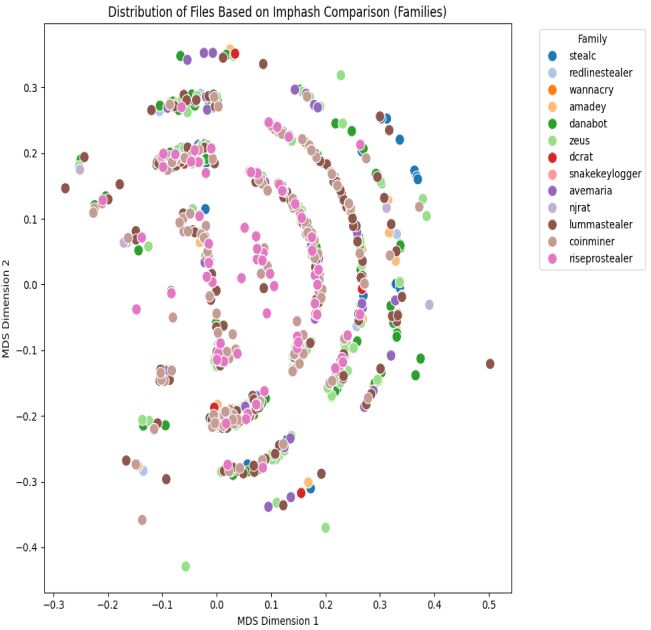}}%
  }
  \caption{Results of methods as families.}
  \label{fig:img_group_results}
  \vspace{-13pt}

\end{figure}





\begin{figure}[t!]
  \centering
  \makebox[\textwidth][c]{%
    \subfloat[Silhouette score of TLSH.]%
      {\includegraphics[width=0.8\textwidth]{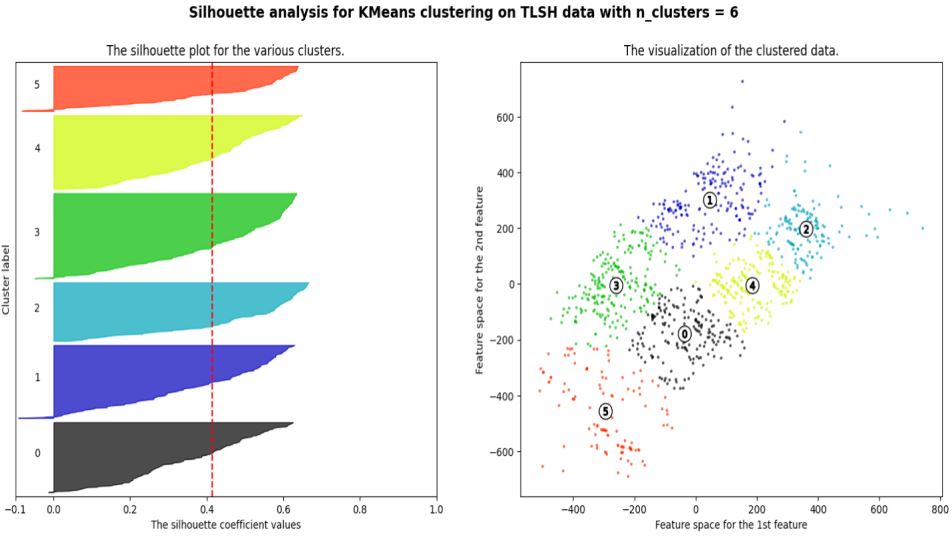}}\hspace{0.8em}%
    \subfloat[Silhouette score of SSDeep.]%
      {\includegraphics[width=0.8\textwidth]{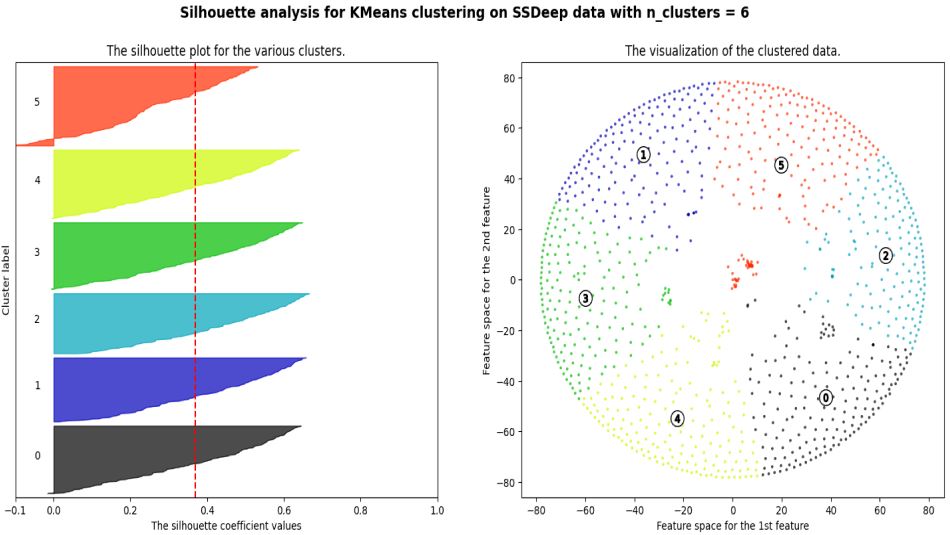}}%
  }
  \caption{Silhouette score of methods.}
  \label{fig:img_group_sil}
  \vspace{-20pt}

\end{figure}

To assess semantic alignment, we overlaid family labels onto the clusters. In the TLSH view (Fig. \ref{fig:img_group_results} (a)), same-family samples form tight, well-separated groups with minimal boundary overlap. SSDeep with family labels (Fig. \ref{fig:img_group_results} (b)) shows that only certain families, such as \textit{njrat} and \textit{wannacry}, stand out as clear clusters, while most others blend together, confirming SSDeep’s strength in broad categorisation but limited fine-grained discrimination. IMPHash (Fig. \ref{fig:img_group_results} (c)) again delivers clean separations, underlining the import-sequence signature’s power to link samples by author or toolchain, even when other file features vary. We further validated cluster quality using the silhouette score over a range of \(K\) values. As shown in Figures \ref{fig:img_group_sil} (a) and (b), both TLSH and SSDeep achieve their highest average silhouette at \(K=6\), indicating that six clusters strike the best balance between cohesion and separation in our dataset. This quantitative result aligns with our qualitative observations, supporting TLSH’s superior fine-grained grouping and confirming SSDeep’s role in coarser partitioning.

  \vspace{-10pt}



\vspace{-5pt}

\section{Conclusions}
\label{conclusion}
\vspace{-10pt}

In this work, we demonstrated that clustering malware using only hash-based features is feasible and efficient. Hashes provide compact, fixed-length representations, enabling fast comparisons for real-time use and integrating seamlessly into existing systems. In our experiments, among the three methods used in our approach, IMPHash performed better than SSDeep and TLSH. The findings also suggest that no single algorithm is universally optimal, since each shows particular strengths depending on the clustering context and the nature of the malware. This highlights the importance of algorithm selection in malware analysis workflows, since understanding these differences can support more effective threat detection and classification, contributing to more robust and adaptive cybersecurity defences.

Potential avenues for future work involve expanding malware datasets by collecting samples across diverse platforms and timeframes to evaluate clustering stability as new variants emerge. Additionally, integration with fuzzy hash functions such as \textit{impfuzzy} and \textit{sdhash} to investigate ensemble-based similarity techniques to improve malware family distinction. To assess adversarial robustness, common obfuscation and evasion techniques could be utilised, including packing, junk insertion and import renaming in order to examine each hash's resilience.

\vspace{-20pt}

\bibliographystyle{splncs04}
\bibliography{dissertationref}

\end{document}